\documentclass[pre,aps,pacs,twocolumn,eqsecnum]{revtex4}
\usepackage{epsfig}
\usepackage{bm}
\usepackage{color}

\newcommand{\C}[1]{{\mathcal{#1}}}
\newcommand{\pa}{\partial}
\usepackage[latin1]{inputenc}
\newcommand{\beq}{\begin{equation}}
\newcommand{\eeq}{\end{equation}}
\newcommand{\bea}{\begin{eqnarray}}
\newcommand{\eea}{\end{eqnarray}}

\begin{document}
\title{Effective Temperature Dynamics in an Athermal Amorphous Plasticity Theory}
\author{Eran Bouchbinder}
\affiliation{Racah Institute of Physics, Hebrew University of Jerusalem, Jerusalem 91904, Israel}
\date{\today}
\begin{abstract}
Recent developments in the theory of amorphous plasticity point to
the central role played by the concept of an effective disorder
temperature $T_{eff}$. An athermal dynamics for $T_{eff}$ are proposed in the framework of a deformation theory and discussed in light of the recent steady state simulations by Haxton and Liu [Phys. Rev. Lett. {\bf 99}, 195701 (2007)]. The structure of the resulting theory, its parameters and transient dynamics are discussed and compared to available data.
\date{\today}
\end{abstract}
\maketitle

\section{Introduction}
\label{int}

Much recent work has been devoted to the detailed experimental, simulational and theoretical analyses of the dynamics of low temperature plasticity in amorphous systems \cite{01LN, 06MR, 07LKXOD, ONO, 04DA, DA1, DA2, 06AD, 07SKLF, 07HL, 07IB, 98FL, 04Lan, 07BLPa, 07BLPb, 07LM}. The systems of interest include noncrystalline
solids well below their glass temperature $T_g$, dense granular
materials, and various kinds of soft materials such as foams,
colloids, and the like. In spite of these efforts there remain fundamental open questions that call for further investigation. Some of the most important emerging new results pointed to the central role played by the concept of an effective disorder temperature $T_{eff}$ \cite{ONO, 04Lan, 07SKLF, 07BLPa, 07BLPb, 07LM, 07MLC, 07HL, 07IB}. It has been proposed that although conventional thermal fluctuations are of little importance for thermal temperatures well below the glass transition temperature $T_g$ or absent in purely athermal systems, the state of configurational disorder of the deforming system can be characterized by an effective disorder temperature $T_{eff}$ that controls configurational fluctuations \cite{04Lan}. However, a well-established equation of motion for $T_{eff}$ is still missing. The aim of the present work is to discuss such an equation of motion and to rationalize its structure based on general considerations. The recent steady state simulations by Haxton and Liu (HL) \cite{07HL} are shown to be consistent with the equation and are used to determine important ingredients of the theory. Predictions for transient dynamics are presented and shown to agree with experimental findings. Finally, open questions are discussed.

There has been some recent discussion in the literature regarding the mathematical rigor, validity and generality of the concept of an effective temperature. It turns out that some measures of an effective temperature agreed numerically with one another, while others did not \cite{TeffREF}. Furthermore, an exact solution of a simple non-equilibrium model shows that there exists a full hierarchy of effective temperatures, that although might agree rather well numerically, have different analytic forms \cite{06SL}. Therefore, it seems that for non-equilibrium systems there are many ways to define a temperature that quantifies deviations from thermal equilibrium, many of which yield numerically consistent values (for reasons that are not yet well understood), but some do not. It is thus natural to ask how can one proceed to develop a theory based on a concept that is not yet completely well-founded?

My answer is that there exists ample evidence for the existence and importance of an intensive macroscopic state variable that characterizes the state of disorder of driven amorphous systems (see below) and therefore one should try to make progress in a phenomenological way based on general considerations and experimental/simulational findings.
In saying so, I do not mean that we should give up trying to develop a rigorous effective temperature formalism. On the contrary, this is a major and fundamental challenge. My main point is that one can make significant progress even in the absence of such an exact formalism.

From a fundamental point of view it is clear that a deformation theory of amorphous systems must include a measure of disorder as a basic dynamical ingredient; structural disorder is an essential feature of these systems. Furthermore, we have a lot of evidence for structural evolution during deformation \cite{07SHR}. We also know that deformation history depends on the system preparation procedure, which is naturally accounted for by assigning different states of initial disorder to different preparation procedures \cite{04DA, DA1, DA2, 06AD, 07SHR}. In addition, there are experiments in which significant structural differences between a material within a shear band and a material outside the band were observed, indicating a structural sensitivity to the rate of deformation (or the rate of energy dissipation) \cite{07SHR}. In fact, there is evidence for structural softening, in addition to thermal softening, within shear bands \cite{07SHR, 05LG}. Recent experiments show that macroscopic quantities like the shear modulus are functions of the deformation \cite{07HDJT}, a point that will be elaborated on later.

An excellent example of these effects was given recently in a series of simulations by Demkowicz and Argon \cite{04DA, DA1, DA2, 06AD}. These authors were able to identify and quantify local structural features of their deforming simulated amorphous silicon. They followed the evolution of these structural measures and demonstrated beautifully the effect of the system preparation procedure on transient dynamics as well as the approach to a unique steady state (independent of initial conditions) under persistent deformation. Similar evidence for the existence of steady state of the disorder was found earlier \cite{ONO}. In my opinion, all this accumulating evidence is pointing towards the need to develop a description of the evolution of disorder in driven amorphous systems and its relation with plastic deformation along physically sensible guidelines, even prior to the availability of an exact non-equilibrium formalism.

The term ``athermal'' is used in this work to refer to situations in which spontaneous thermal fluctuations are incapable of inducing structural rearrangements. Therefore, in these situations plastic deformation occurs only in response to external driving forces. For example, such conditions are relevant for glasses well below their glass transition temperature or for granular media and foams where thermal fluctuations are practically nonexistent. By using this term we by no means imply that thermal vibrations or dissipative mechanisms like friction in granular media or viscosity in foams are not important; on the contrary these processes are crucial for our discussion below, where they provide a means to remove energy irreversibly from the deforming system once structural rearrangements induced by the external driving forces take place. The ideas developed hereafter are presented in a way that makes explicit reference to vibrational motion. However, the resulting framework applies equally well to systems where thermal vibrations are completely absent.

In the next section an athermal dynamics for $T_{eff}$ are proposed and discussed in relation to a deformation theory. The resulting dynamics are compared to the steady state Haxton and Liu (HL) data \cite{07HL} in Sect. \ref{simulation}. In Sect. \ref{transient} we go beyond the steady state analysis to predict some transient effects. Section \ref{sum} offers a summary and some discussion.

\section{Athermal $T_{eff}$ dynamics and a deformation theory}
\label{dynamics}

The basic starting point for the discussion of the effective temperature dynamics is the separation of the total amount of degrees of freedom of the system under consideration to fast vibrational and slow configurational ones. This separation, known as the ``inherent states'' formalism \cite{Inherent}, was shown to provide a proper framework to describe glassy dynamics; the phase space dynamics are such that the system remains in the basin of a single minimum in the potential energy landscape (an ``inherent state'' or a ``configuration'') for a long time compared to the particles vibration timescale until it makes a transition to another basin \cite{Inherent, 00BBK}. Our goal here is to extend these ideas to dissipative, driven (out of mechanical equilibrium) amorphous systems.

In the athermal limit considered in this work the transitions between the ``inherent states'' occur only as a result of the applied forces and not due to spontaneous thermal fluctuations that are assumed to be inefficient at low enough temperatures. The response of such systems to the application of external driving forces typically contains a reversible (elastic) and an irreversible (plastic) components. The plastic component, usually termed the plastic part of the rate of deformation tensor, is denoted by $\hat{D}^{pl}$. The plastic work density per unit time is $s_{ij}D^{pl}_{ij}$, where $s_{ij}$ is the deviatoric part of the stress tensor. At low temperatures, well below the glass transition one, non of this energy can be stored reversibly, implying that $s_{ij}D^{pl}_{ij}$ is all dissipated \cite{07BLPa}.

In light of the separation of degrees of freedom we can write the total heat flux $\dot{Q}$ as
\begin{eqnarray}
\label{total_heat}
\dot{Q}=\dot{Q}_v+\dot{Q}_c \ ,
\end{eqnarray}
where $v$ and $c$ stand for vibrational and configurational respectively. Generally speaking, $\dot{Q}_v$ is the energy flux that is being removed irreversibly from the deforming system to its surroundings. In atomic or molecular systems at a finite $T$, it is being transferred to the heat bath by thermal vibrations. In systems where thermal vibrations are absent, some other physical mechanism like internal friction in granular media or viscosity in foams, is responsible for irreversibly removing energy from the deforming system. $\dot{Q}_c$ is the rate of change of the energy density stored in the configurational degrees of freedom.
The energy dissipation rate equals the total heat flux, therefore
\begin{eqnarray}
\label{dissipation}
s_{ij}D^{pl}_{ij} = \dot{Q} \ .
\end{eqnarray}
Following Langer \cite{04Lan} we propose that the configurational degrees of freedom are characterized by an effective disorder temperature $T_{eff}$, possibly different from the bath temperature $T$, such that
\begin{eqnarray}
\label{config_heat}
c_{eff}\dot{T}_{eff} = \dot{Q}_c \ ,
\end{eqnarray}
where $c_{eff}$ is the configurational specific heat of units $k_B$ per unit volume. Using Eqs. (\ref{total_heat})-(\ref{Teff_dot}) we obtain
\begin{eqnarray}
\label{Teff_dot}
c_{eff}\dot{T}_{eff} = s_{ij}D^{pl}_{ij}-\dot{Q}_v \ .
\end{eqnarray}
The quantity $\dot{Q}_v$, as explained above, represents the rate at which energy is removed irreversibly from the deforming system to the heat bath. In general, a first principles calculation that accounts for the detailed processes that involve heat exchanges with the bath is well beyond the scope of the present work and is anyway an extremely difficult task. However, the athermal limit implies some simplifications. In that limit, $\dot{Q}_v$ vanishes in the absence of plastic deformations, i.e. when $D^{pl}_{ij}\!=\!0$. Moreover, $\dot{Q}_v$ is a scalar that is related to dissipation and has the dimension of energy density per unit time. These properties are naturally accounted for by assuming that $\dot{Q}_v$ is proportional to the plastic power density $s_{ij}D^{pl}_{ij}$, i.e. that
\begin{eqnarray}
\label{vib_heat}
\dot{Q}_v = s_{ij}D^{pl}_{ij}g(s_{ij},T_{eff},T) \ ,
\end{eqnarray}
where $g(s_{ij},T_{eff},T)$ is a dimensionless function.
This function has a clear physical meaning as it represents the fraction of the plastic power density $s_{ij}D^{pl}_{ij}$ that is being transformed into regular heat. For example, when it equals unity all the plastic power density $s_{ij}D^{pl}_{ij}$ is converted into regular heat and non of it is being stored in the configurational degrees of freedom.
Substituting Eq. (\ref{vib_heat}) into Eq. (\ref{Teff_dot}) we obtain
\begin{eqnarray}
\label{Teff_dot1}
c_{eff}\dot{T}_{eff} = s_{ij}D^{pl}_{ij}\left[1-g(s_{ij},T_{eff},T)\right] \ .
\end{eqnarray}
The structure of this equation is consistent with the original equation proposed by Langer \cite{04Lan}.
Following these ideas, we can write the total entropy density $S$ as
\begin{eqnarray}
\label{entropy}
S=S_v+S_c \ ,
\end{eqnarray}
and identify
\begin{eqnarray}
\label{entropy_heat}
\dot{Q}_c = T_{eff} \dot{S}_c \ .
\end{eqnarray}
The last relation implies that
\begin{equation}
\label{Teff_der}
\frac{1}{T_{eff}}=\frac{\pa S_c}{\pa E_c} \ ,
\end{equation}
where $E_c$ is the configurational energy density (``inherent states'' energy density). Eq. (\ref{Teff_der}) can serve as a definition of the effective temperature (also called the configurational temperature) and in fact was used in relation to Edwards' hypothesis \cite{Edwards} in Ref. \cite{02CFN}, where it was shown that a generalized statistical mechanics for the ``inherent states'' can be developed. More important to our discussion here are the results of \cite{ONO, 02MK} that show that Eq. (\ref{Teff_der}) yields values of $T_{eff}$ that are consistent with standard thermodynamic relations applied to out of equilibrium systems. Note also that the above relations imply that for out of equilibrium situations, i.e. when $T_{eff}\!\ne\!T$, we have
\begin{equation}
\dot{Q} = T \dot{S}_v+T_{eff} \dot{S}_c\ne T\dot{S} \ ,
\end{equation}
which is an expected violation of a standard thermodynamic relation.

The development up to now focused on a quasi-thermodynamic interpretation of the effective temperature $T_{eff}$. In this spirit, it is natural to ask whether gradients of $T_{eff}$ drive a configurational heat flux from high to low $T_{eff}$ regions, in analogy with a regular temperature behavior. Alternatively, we can ask whether a diffusion term of the form $D_{eff}\nabla^2T_{eff}$ should be added to the right-hand-side of Eq. (\ref{Teff_dot1}), where $D_{eff}$ is an effective diffusion coefficient. From the discussion above it is clear that if $T_{eff}$ diffuses, it must be associated with plastic flow, i.e. with $D^{pl}\!\ne\!0$. Therefore, dimensional analysis suggests that $D_{eff}\!\propto\!D^{pl}$. The most natural phenomenon in which this behavior should be observed is shear banding. When shear banding occurs, we expect a high $T_{eff}$ region (inside the band, where $D^{pl}$ is large) to coexist with a low $T_{eff}$ region (outside the band, where $D^{pl}$ is small), see for example \cite{07MLC}. Indeed, the shear banding simulations of \cite{07SKLF} suggest that the width of the shear band broadens diffusively, i.e. as a
function of strain (or time) to the power $1/2$ \cite{Falk}. This also implies that the rate of broadening is
proportional to $D^{pl}$ in the band. A systematic analysis of such effects is needed in order to provide further support for the diffusivity of $T_{eff}$. In this work we focus on homogeneous situations, such that we can omit a possible diffusion term in Eq. (\ref{Teff_dot1}).

Eq. (\ref{Teff_dot1}) becomes useful only if $D^{pl}_{ij}$ and $g(s_{ij},T_{eff},T)$ are known. We first discuss $D^{pl}_{ij}$ in the framework of the recently proposed Shear-Transformation-Zones (STZ) theory \cite{98FL,04Lan, 07BLPa}. Building on early ``flow-defect'' ideas \cite{EarlySTZ}, this theory views
the plastic deformation of an amorphous solid as consisting of finite strain, localized, irreversible rearrangements that take place in response to applied forces. A major theoretical challenge is the calculation of the probability of occurrence of these these irreversible ``events'', the so-called shear transformations, or more precisely the rate of events which gives rise to a macroscopic plastic rate of deformation $D^{pl}_{ij}$.
The STZ theory culminates in the following expression for $D^{pl}_{ij}$ \cite{04Lan, 07BLPa}
\begin{equation}
\label{Dpl}
D^{pl}_{ij}=\exp{\left(-\frac{G_{STZ}}{k_B T_{eff}}\right)} \frac{f_{ij}(s_{ij})}{\tau_0} \ ,
\end{equation}
which is valid in the absence of extremely fast transients \cite{07BLPb}. The validity of this form was demonstrated recently in a shear banding simulation \cite{07SKLF}.

Let us discuss the physical meaning of this expression. The Boltzmann-like factor constitutes the configurational part of $D^{pl}_{ij}$, a part that is proportional to the probability to find a local configurational fluctuation that might enable a plastic event to take place. More explicitly, in light of Eq. (\ref{Teff_der}) and in the spirit of Ref. \cite{02CFN}, we postulate that the probability to observe a free energy fluctuation $\delta G > G_{STZ}$ is given by
\begin{equation}
\label{distribution}
p(\delta G) \propto  \frac{1}{k_B T_{eff}}\exp{\left(-\frac{\delta G}{k_B T_{eff}}\right)} \quad\hbox{for}\quad \delta G >G_{STZ}\ .
\end{equation}
The free energy $G_{STZ}$ represents the existence of a minimal free energy fluctuation below which a shear transformation cannot occur at all, making the cumulative probability to find a potential site for a plastic event proportional to
\begin{equation}
\int_{G_{STZ}}^\infty p(\delta G) d(\delta G) =  \exp{\left(-\frac{G_{STZ}}{k_B T_{eff}}\right)} \ ,
\end{equation}
as appears in Eq. (\ref{Dpl}).

The additional multiplicative term $f_{ij}(s_{ij})/\tau_0$ represents that fact that the existence of a potential site for a plastic event does not immediately imply the occurrence of an event. There are also dynamic factors that determine the actual rate of irreversible events. First, the time scale $\tau_0$ determines the basic attempt period and was proposed to be of the order of the vibration time scale for atomic systems \cite{07BLPb, 07LM, 07MLC}. For other systems, this time scale is the characteristic relaxation time; such a time scale always exists, even when thermal fluctuations are absent. The other factor $f_{ij}(s_{ij})$ quantifies the explicit dependence of $\hat{D}^{pl}$ on the deviatoric stress $\hat{s}$. The existence of such a term is anticipated on the basis of symmetry as $\hat{D}^{pl}$ is a tensorial quantity. In fact, STZ theory offers a more detailed description of the tensorial function $f_{ij}(s_{ij})$, including an additional internal state tensor field that accounts for the transition from jamming to flow at the yield stress and for orientational memory \cite{98FL,07BLPa,07BL}. In summary, as described by Eq. (\ref{Dpl}), STZ theory provides an expression for the rate of irreversible (plastic) events, incorporating both configurational and dynamic factors.

The plastic rate of deformation tensor $\hat{D}^{pl}$ is incorporated into an elaso-plastic deformation theory by assuming that the total rate of deformation tensor $\hat{D}^{tot}$ can be written as a simple sum of linear elastic and plastic contributions
\begin{equation}
\label{Dtot}
D^{tot}_{ij}=\dot{\epsilon}^{el}_{ij}+D^{pl}_{ij} \ ,
\end{equation}
where $\epsilon^{el}_{ij}$ is the linear elastic strain tensor \cite{86LL}. This equation, upon substituting $D^{pl}_{ij}$ of Eq. (\ref{Dpl}), together with Eq. (\ref{Teff_dot1}), constitute a compact deformation theory. In order to complete the derivation of the theory one must specify the functions $g(s_{ij},T_{eff},T)$ and $f_{ij}(s_{ij})$. As a first principles derivation of these functions is not yet available, we have to resort to molecular dynamics simulations.

\section{The Haxton and Liu data}
\label{simulation}

An information of the type we are looking for became available recently through the remarkable steady state simulations by Haxton and Liu (HL)
\cite{07HL}. These authors performed molecular dynamics simulations of a
purely repulsive binary glass, made of equal proportion of small and
large particles and sheared steadily under a fixed area
fraction. We write the tensors in the problem as
\begin{equation}
\hat{s}=\left(\begin{array}{cc}0&s\\s&0\end{array}\right),\quad
\hat{D}^{pl}=\left(\begin{array}{cc}0&D^{pl}\\D^{pl}&0\end{array}\right) \ ,
\end{equation}
and similarly for $\hat{D}^{tot}$; we stress the fact that there is only one independent component in each one of them, making the problem effectively scalar. Note also that under fixed area the pressure is expected to vary with deformation. Modeling direct pressure effects, for example the dependence of the yield stress on the pressure, requires additional approximations and assumptions that are not essential for the present discussion. We therefore keep adopting the assumption made above that plastic deformation is mainly coupled to the deviatoric part of the stress and neglect additional pressure effects.

The ordinary thermal
temperature $T$ in the HL simulations was varied systematically from well above the glass
temperature to well below and three orders of magnitude of imposed
total rate of deformation $D^{tot}$ were studied, ranging from
$\tau_0 D^{tot}\!=\!10^{-5}$ to $\tau_0D^{tot}\!=\!10^{-2}$, where $\tau_0$ is
roughly the binary collision time. The effective temperature was
evaluated using a relation between the static linear
response and the variance of the pressure. The crucial point is that this measurement of $T_{eff}$ was shown in \cite{ONO} to yield values that are numerically consistent with our definition of Eq. (\ref{Teff_der}), making HL data most relevant for the developed formalism.

Since our main interest here is in the athermal limit, we focus on the low temperatures data and take the
$T\!\approx\!0.2T_0$ results, where $T_0$ is estimated from a Vogel-Fulcher
fit of the equilibrium viscosity, as a representative example. All the results are given in units of $\tau_0$, the
particle mass, the small particles diameter, the interaction
energy scale and $k_B\!=\!1$. It is important to note that as the total rate of deformation $D^{tot}$ was controlled in these numerical experiments and only steady states were considered, the deviatoric stress $s$ is always larger than the yield stress $s_y$, which is defined as
\begin{equation}
s_y \equiv \lim_{q\to 0} s \ ,
\end{equation}
where $q\!\equiv\!\tau_0D^{tot}$.

\begin{figure}
\centering \epsfig{width=.5\textwidth,file=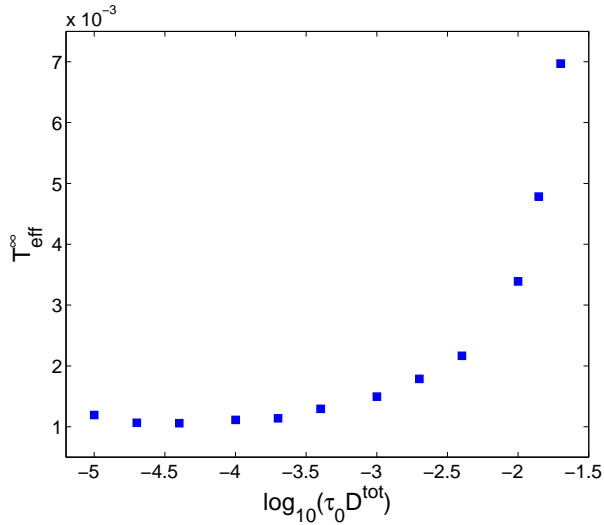}
\caption{$T^\infty_{eff}$ vs. $log_{10}\left(\tau_0D^{tot}\right)$ in the simulations of Ref. \cite{07HL}.
Data courtesy of T. Haxton and A. Liu.}\label{HL}
\end{figure}

The steady state deviatoric stress $s^\infty\!>\!s_y$, as well as the steady state effective temperature
$T^\infty_{eff}$, were measured as a function of the normalized imposed total rate of deformation $\tau_0D^{tot}$. A relation between the two was found to be of the form
\begin{equation}
\label{steady_s} s^\infty \approx s_0 \exp{\left(-\frac{G_r}{T^\infty_{eff}}\right)} \ ,
\end{equation}
where $G_r\!\approx \!0.0028$ and $s_0\!\approx\!0.011$ are
independent of both $T$ (for $T\!\ll\!T_0$) and $\tau_0 D^{tot}$.
$T^\infty_{eff}$ itself is plotted in Fig. \ref{HL} as a function of $\tau_0D^{tot}$. It is observed that $T^\infty_{eff}$ approaches a limiting value in the small $\tau_0D^{tot}$ limit and exhibits a marked increase for $\tau_0D^{tot}\!>\!10^{-3}$, an effect that was not taken into account in previous works \cite{04Lan,07BLPa,07MLC}.

These results are completely consistent with the $T_{eff}$ dynamics proposed in Eq. (\ref{Teff_dot1}). The steady state relation of Eq. (\ref{steady_s}) can be recovered, in the framework of the proposed theory, by choosing (putting back $k_B$)
\begin{equation}
\label{g} g(s_{ij},T_{eff}, T) \approx \frac{s_0}{\bar{s}}\exp{\left(-\frac{G_r}{k_B T_{eff}}\right)}\quad\hbox{for}\quad \bar{s}>s_y \ ,
\end{equation}
with
\begin{equation}
\label{bar_s}
\bar{s}\equiv\sqrt{\frac{s_{ij}s_{ij}}{2}} \ .
\end{equation}
Substituting Eq. (\ref{g}) into Eq. (\ref{vib_heat}) we obtain
\begin{eqnarray}
\label{vib_heat1}
\dot{Q}_v = s_{ij}D^{pl}_{ij}\frac{s_0}{\bar{s}}\exp{\left(-\frac{G_r}{k_B T_{eff}}\right)}\quad\hbox{for}\quad \bar{s}>s_y \ .
\end{eqnarray}

Let us try to interpret these last few results. The structure of $\dot{Q}_v$, i.e. the rate at which energy is being removed irreversibly from the deforming system to its surroundings, is suggestive. The plastic power $s_{ij}D^{pl}_{ij}$ part was discussed above. The remaining multiplicative factor, which is the function $g$ of Eq. (\ref{g}), represents the fraction of that power that flows to the heat bath as regular heat. It depends both on $T_{eff}$ and on the deviatoric stress through $\bar{s}$. The appearance of $T_{eff}$ in an exponential Boltzmann-like factor gives us some confidence that Eq. (\ref{vib_heat1}) presents a physically meaningful result and not merely a fudge factor, in light of role played by $T_{eff}$ in the proposed theory.
This exponential Boltzmann-like factor must be related to the distribution of inherent states that determines the probability of increasing/decreasing the energy stored in the configurational degrees of freedom during irreversible plastic rearrangements, though additional theoretical work is needed in order to derive it from first principles. An important feature of Eq. (\ref{g}) is that the fraction of $s_{ij}D^{pl}_{ij}$ that is converted to regular heat is an increasing function of $T_{eff}$. However, contrary to previous works \cite{04Lan, 07BLPa, 07LM, 07MLC}, $g$ also depends explicitly on the stress through $\bar{s}$. This dependence is such that $g$ decreases with increasing $\bar{s}$, if we think of $T_{eff}$ as fixed (which is not the case). The fact that $g$ in Eq. (\ref{g}) depends both on $T_{eff}$ and $\bar{s}$ opens the interesting possibility, as will be shown below,  that during transient dynamics energy will be removed temporarily from the configurational degrees of freedom, resulting in a decrease in $T_{eff}$. That possibility corresponds to $g(s_{ij},T_{eff})$ in Eq. (\ref{Teff_dot1}) being larger than unity. This effect was demonstrated experimentally in \cite{99Rittel} and will be discussed in detail in Sect. \ref{transient}.

In passing we note that Eq. (\ref{Teff_dot1}) must be invariant under the symmetry operations $s_{ij}\!\to\!-s_{ij}$ and $D^{pl}_{ij}\!\to\!-D^{pl}_{ij}$.
Therefore, the function $g(\bar{s},T_{eff})$ in Eq. (\ref{g}) (which is strictly valid for $\bar{s}\!>\!s_y$) should be smoothly interpolated towards $\bar{s}\!=\!0$, with a continuous derivative there, for example using $s_0/\sqrt{\bar{s}^2+s_1^2}$ with $s_1\!\ll\!s_y$. In spite of the fact that the steady state data for $\bar{s}\!>\!s_y$, corresponding to a finite $D^{tot}$, cannot determine precisely the behavior for $\bar{s}\!<\!s_y$ we stress the fact that $g$ is an analytic function that shows no special behavior near $\bar{s}\!\to\!0$. In this particular relation it is important to note that in interpreting their data HL proposed, following the free-volume theory of \cite{02Lemaitre}, an equation for $T_{eff}$ that is consistent with the structure of our Eq. (\ref{Teff_dot1}). Specifically, for the relevant configuration they proposed (using our notation and terminology) that
\begin{equation}
\label{gHL} \dot{Q}_v \approx 2 s_0 D^{pl}\exp{\left(-\frac{G_r}{k_B T_{eff}}\right)} \ ,
\end{equation}
which is indeed identical to Eq. (\ref{vib_heat1}) for $s\!>\!s_y$.
However, the tensorial generalization of the last relation involves replacing $D^{pl}$ with $\bar{D}^{pl}$, which is defined in complete analogy with Eq. (\ref{bar_s}). This generalization exhibits a non-analytic behavior in the limit $\bar{D}^{pl}\!\to\!0$. This kind of non-analyticity was criticized in \cite{04FLP} as being unlikely to arise from any first principles analysis of molecular mechanisms. We therefore prefer not to incorporate non-analytic behavior in our theory as long as we are not forced to.

The information about the steady state effective temperature shown in Fig. \ref{HL} should be now used to extract the only missing piece in our theory, i.e. the tensorial function $f_{ij}$ in Eq. (\ref{Dpl}). It has the following structure within the STZ theory \cite{07MLC, 07BLPS}
\begin{equation}
\label{fSTZ}
f_{ij}(s_{ij},m_{ij})=\overline{\C C}(\bar{s})\left(\frac{s_{ij}}{\bar{s}}-m_{ij}\right) \ ,
\end{equation}
where $\overline{\C C}(\bar{s})$ is a dimensionless scalar rate function and $\hat{m}$ is an internal state tensor field that represents the orientational properties of the plastic events and accounts for memory effects and the transition from jamming to flow at the yield stress $s_y$ \cite{98FL, 07BLPa, 07BL}.

Let us discuss the physical origin of the tensorial function $f_{ij}(s_{ij},m_{ij})$. The exponential Boltzmann-like factor (that depends on $T_{eff}$) appearing in Eq. (\ref{Dpl}) is proportional to the average number of available sites for plastic rearrangements. The actual rate of plastic rearrangements, which determines the macroscopic plastic rate of deformation $\hat{D}^{pl}$, depends {\em also} on the current deviatoric stress and the recent history of deformation as quantified by $f_{ij}(s_{ij},m_{ij})$. The internal state field $\hat{m}$ quantifies the effect of recent history on the deformation and for our purposes here should be thought of as a normalized back stress. That means that after a plastic event in a certain direction took place, further deformation in that direction is more difficult, which is represented by the term $-m_{ij}$ in Eq. (\ref{fSTZ}) \cite{07BLPa}. Therefore, the effective stress at a given point is directed according to $s_{ij}/\bar{s}-m_{ij}$. This expression contains all the tensorial information about the symmetry of the deformation, representing its direction (but not its magnitude). In an isotropic material, the deformation should follow the symmetry of the local stress $s_{ij}$; however, recent deformation carries some orientational memory that is represented by $m_{ij}$ whose principle axes do not necessarily coincide with those of $s_{ij}$.

The remaining multiplicative factor in Eq. (\ref{fSTZ}), $\overline{\C C}(\bar{s})$, represents the effect of the magnitude of the stress on the rate of actual plastic events. This function is certainly the most phenomenological part of the STZ theory. A first principles derivation of such a material dependent function is probably unrealistic and it seems that the best approach to study it is by microscopic simulations. In the present context we only aim at showing that there exists a function $\overline{\C C}(\bar{s})$, with the general appropriate features, that can describe properly the numerical data of HL. First, we expect $\overline{\C C}(\bar{s})$ to be a monotonically increasing function of $\bar{s}$. Furthermore, following \cite{07LM}, we expect that for small stresses the dynamics of plastic rearrangements involve barrier crossings between different inherent states in the energy landscape. Therefore, for $\bar{s}$ of the order of $s_y$ or less, we expect activated dynamics to dominate, resulting in $\overline{\C C}(\bar{s})$ that is at least an exponential function of $\bar{s}$. For higher stresses, we expect the notion of barriers to be irrelevant and the rate of plastic rearrangements should depend on some other dissipative mechanism, resulting in a weaker dependence of $\overline{\C C}(\bar{s})$ on $\bar{s}$. In addition, $\overline{\C C}(\bar{s})$ is a symmetric function that vanishes for $\bar{s}\!=\!0$ in the athermal limit, representing the fact that there cannot be irreversible events in a direction opposite to the applied force in the absence of efficient thermal fluctuations \cite{07BLPa}. This property is not relevant for the HL data since they report on steady state results with $\bar{s}\!>\!s_y$.

To prepare the equations for the comparison with the HL data, we note that in a simple shear configuration the fixed-points of the only non-vanishing independent component of $\hat{m}$ are \cite{07BLPb}
\begin{equation}
\label{m0} m_0(s)=\cases{s/|s| &if $|
s|\le s_y$\cr s_y/s & if $|s| > s_y$\ .}
\end{equation}
In \cite{07BLPb} it was shown that $m$ relaxes to one of its fixed-points on a time scale much smaller than the typical time scale of the $T_{eff}$ dynamics, enabling us to replace the $m$-dynamics by one of its fixed-point.
Therefore, for steady state conditions under simple shear, Eq. (\ref{fSTZ}) simplifies to
\begin{equation}
\label{fSTZ1}
f(s)=\overline{\C C}(s)\left(\frac{s}{|s|}-m_0(s)\right) \ .
\end{equation}
We thus conclude the for the purpose of our analysis of the HL data, we can use the $s\!>\!s_y$ branch in Eq. (\ref{m0}) such that Eq. (\ref{Dpl}) reduces to
\begin{equation}
\label{DplSS}
D^{pl}=\exp{\left(-\frac{G_{STZ}}{k_B T_{eff}}\right)}\frac{\overline{\C C}(s)}{\tau_0}\left(1-\frac{s_y}{s}\right) \ .
\end{equation}

\begin{figure}
\centering \epsfig{width=.5\textwidth,file=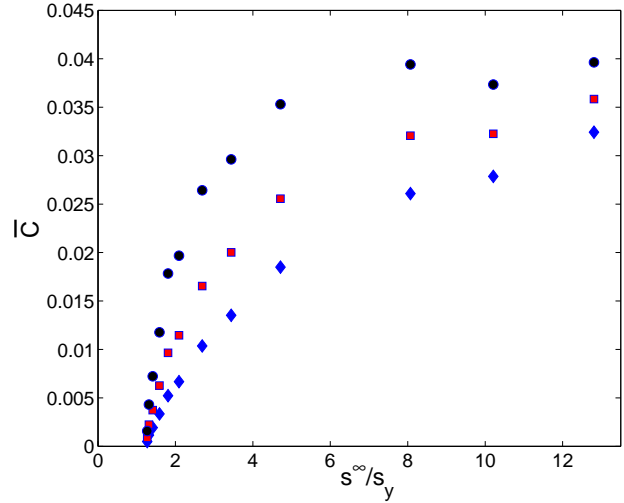}
\caption{$\overline{\C C}$ as a function $s^\infty/s_y$ for $\beta\!=\!1$ (diamonds), $\beta\!=\!1.25$ (squares) and $\beta\!=\!1.5$ (circles).}\label{C_s}
\end{figure}

We now proceed to discuss the estimation of $\overline{\C C}(s)$. For that aim we define $\beta\!\equiv\! G_{STZ}/G_r$ and then substitute Eq. (\ref{steady_s}) into Eq. (\ref{DplSS}) to obtain
\begin{equation}
\label{extractingCs}
\overline{\C C}=\frac{\tau_0 D^{tot} \left(\displaystyle\frac{s^\infty}{s_0} \right)^\beta}{1-\displaystyle\frac{s_y}{s^\infty}} \ ,
\end{equation}
where $D^{tot}$ is considered as a function of $s^\infty$. From Eq. (\ref{extractingCs}) we learn that $\overline{\C C}$ depends both on the yield stress $s_y$ and $\beta$. Estimating $s_y$ from the data is a rather subtle issue as it involves an extrapolation toward the $\tau_0D^{tot}\!\to\!0$ limit. However, for the sake of our discussion, the exact value is not crucial and we choose to estimate $s_y$ from Eq. (\ref{steady_s}) with $T^\infty_{eff}(\tau_0D^{tot}\!\to\!0)\!\simeq\!0.001$ as the low strain rate limit in Fig. \ref{HL}. The parameter $\beta$, on the other hand, cannot be calculated independently, implying that $\overline{\C C}$ is not uniquely determined by the data. However, we do expect it to be $\C O(1)$ as it is unlikely that two dramatically different energy scales enter the problem. More interestingly, it turns out that a physical constraint bounds $\beta$ from above. The idea is that, as explained above, the transitions rate function $\overline{\C C}$ is a non-decreasing function of the stress $s$ since there cannot be less irreversible events in the direction of the applied force as its magnitude increases. In Fig. \ref{C_s} $\overline{\C C}$ is plotted as function of $s^\infty/s_y$ for $\beta\!=\!1,1.25,1.5$ using Eq. (\ref{extractingCs}).
It is observed that for $\beta\!=\!1.5$ the curve becomes non-increasing. Therefore, $\beta$ is constrained to be smaller than about $1.5$, which seems consistent with our original expectation. If indeed $\beta\!\sim\!\C O(1)$, then the rate function $\overline{\C C}$ is in the range shown in Fig. \ref{C_s}. Note, however, that we cannot derive a strict lower bound for $\beta$. Note also that $\overline{\C C}$ should be properly interpolated such that $\overline{\C C}\!\to\!0$ for $s\!\to\!0$, as required in the athermal limit.

The results presented in Fig. (\ref{C_s}) follow our general expectations regarding the generic features of the material dependent function $\overline{\C C}(s)$. It is observed that it is a monotonically increasing function of the stress and that this dependence is stronger for stresses in the vicinity of $s_y$ than for higher stresses. We conclude that the numerical data of HL is completely consistent with the proposed theory and tightly constrain those parts of it that at the moment cannot be derived directly. In the next section we use the resulting theory to predict transient dynamics that go beyond the steady state data reported in \cite{07HL}. An important feature of our theoretical structure is that the steady state of $T_{eff}$ is {\em not} obtained directly using Eq. (\ref{Teff_dot1}), which with Eq. (\ref{g}), provides only a relation between $s^\infty$ and $T^\infty_{eff}$. The value of $T^\infty_{eff}$ is obtained only after $s^\infty(T^\infty_{eff})$ is substituted in Eq. (\ref{DplSS}). This implies that $\overline{\C C}(s)$ in Eq. (\ref{DplSS}) is the fundamental quantity that determines $T^\infty_{eff}$. Therefore, additional efforts are needed in order to gain a deeper theoretical understanding of this function.

\section{Transient dynamics}
\label{transient}

The analysis of the previous section was based on steady state information alone. In this section we demonstrate some generic new transient features of the proposed $T_{eff}$ dynamics. For that aim we reduce Eq. (\ref{Dtot}) to its simple shear form and use the linear elastic relation $\epsilon^{el}=s/2\mu$, where $\mu$ is the shear modulus, to obtain
\begin{equation}
\label{dot_s}
\dot{s}=2\mu\left[D^{tot}-D^{pl}(s,T_{eff})\right] \ ,
\end{equation}
with
\begin{equation}
\tau_0 D^{pl}(s,T_{eff})=\exp{\left(-\frac{G_{STZ}}{k_B T_{eff}}\right)}\overline{\C C}(s)\left(\frac{s}{|s|}-m_0(s)\right) \ .
\end{equation}
Note that Eq. (\ref{fSTZ1}) was used to allow for arbitrary values of $s$.
For simple shear, Eq. (\ref{Teff_dot1}) (with Eq. (\ref{g}) and a proper interpolation for $|s|\!<\!s_y$) reads
\begin{equation}
\label{heat1}
c_{eff}\dot{T}_{eff}\!=\!\left[2sD^{pl}-\frac{2sD^{pl} s_0}{\sqrt{s^2+s_1^2}}\exp{\left(-\frac{G_r}{k_BT_{eff}}\right)} \right] \ ,
\end{equation}
with $s_1\!\ll\!s_y$.
\begin{figure}
\centering \epsfig{width=.5\textwidth,file=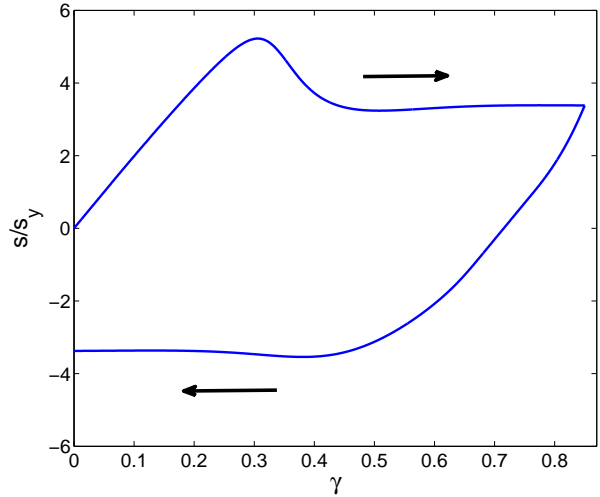}
\caption{The deviatoric stress $s/s_y$ as a function of the imposed strain $\gamma\!\equiv\!\int_0^tD^{tot}(t')dt'$. The arrows indicate the direction of the applied $D^{tot}$.}\label{cycle_s}
\end{figure}

For the sake of illustration we keep using the values of $s_0$, $s_y$ and $G_r$ corresponding to the HL data (in HL units), as well as $G_{STZ}\!=\!2G_r$, $\mu=10s_y$ and $c_{eff}\!=\!1/3$. Note that the units of $c_{eff}$ are $k_B$ per unit volume. In HL units, where $k_B\!=\!1$ and the basic volume is an atomic one, $c_{eff}$ is expected to be of order unity. In addition, we choose
\begin{equation}
\overline{\C C}\!=\!0.001|s/s_y|^6/(1+|s/s_y|^5) \ ,
\end{equation}
which is a simple function with the proper qualitative features discussed above. Note that using instead an interpolated version of one of the curves in Fig. \ref{C_s} would not affect any of the results to follow. We set the initial effective temperature to $T_{eff}(t\!=\!0)\!=\!9.5\times10^{-4}$, a quantity that, for example, can be controlled by varying the rate of cooling by which the glass transition is approached, see \cite{06AD} for details. We analyzed Eqs. (\ref{dot_s})-(\ref{heat1}) for a loading cycle in which $\tau_0D^{tot}\!=\!10^{-4}$ for the first half cycle and $\tau_0D^{tot}\!=\!-10^{-4}$ for the second half. Figure \ref{cycle_s} shows the normalized deviatoric stress $s/s_y$ as a function of the external strain $\gamma\!\equiv\!\int_0^tD^{tot}(t')dt'$. The figure exhibits a rather generic behavior where there exists a stress peak followed by strain softening in the first half cycle and a softer response in the second half cycle due to memory effects associated with $T_{eff}$ \cite{07BLPa}.
\begin{figure}
\centering \epsfig{width=.5\textwidth,file=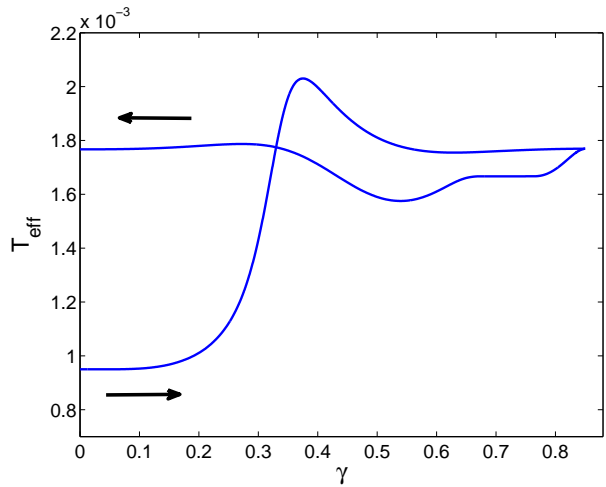}
\caption{The effective disorder temperature $T_{eff}$ as a function of the imposed strain $\gamma\!\equiv\!\int_0^tD^{tot}(t')dt'$. The arrows indicate the direction of the applied $D^{tot}$.}\label{cycle_Teff}
\end{figure}

More interestingly, Fig. \ref{cycle_Teff} shows the corresponding plot for the effective disorder temperature $T_{eff}$ as a function of the imposed strain $\gamma$. The qualitative new feature observed is that during transients $T_{eff}$ might exhibit a {\em non-monotonic} behavior. During the first half cycle $T_{eff}$ overshoots the steady state value $T^\infty_{eff}$, while during the second half cycle it undershoots $T^\infty_{eff}$.
This effect becomes more pronounced as the imposed rate of deformation is increased.

The physical origin of the non-monotonic behavior is the dependence of $g$ in Eq. (\ref{g}) on the stress $\bar{s}$, in addition to $T_{eff}$. If $g$ were a function of $T_{eff}$ alone, then for all cases in which the initial value of $T_{eff}$ is smaller than the steady state one, Eq. (\ref{Teff_dot1}) would predict a monotonic increase of $T_{eff}$, irrespective of the dynamics of the stress $\bar{s}$. However, the dependence of $g$ in Eq. (\ref{g}) on the stress $\bar{s}$, allows for a non-monotonic behavior of $T_{eff}$ during transients. For example, during the first half of the loading cycle discussed above, $g(\bar{s}, T_{eff})$ becomes unity while the stress is still decreasing in the strain softening part observed in Fig. \ref{cycle_s}. That implies that $g(\bar{s}, T_{eff})$ in Eq. (\ref{g}) temporarily overshoots unity such that $\dot{T}_{eff}$ becomes {\em negative}. This behavior suggests a somewhat surprising physical phenomenon in which during deformation transients energy stored in the configurational degrees of freedom is released, i.e. $\dot{Q}_c\!<\!0$, and converted into regular heat such that the amount of heat being removed from the system to its surroundings is {\em larger} than the plastic power $s_{ij}D^{pl}_{ij}$.

The predicted non-monotonic behavior of $T_{eff}$ during transients can be tested directly in laboratory experiments. For that aim, consider the heat equation for the regular temperature $T$
\begin{equation}
\label{heat_equationT}
c\dot{T}=g(\bar{s}, T_{eff}) s_{ij}D^{pl}_{ij} + D \nabla^2 T \ ,
\end{equation}
where $D$ is the thermal diffusion coefficient and $c$ is the thermal specific heat of units $k_B$ per unit volume. Note that thermoelastic effects were neglected here. Under sufficiently fast deformation the heat conduction term can be neglected, leading to adiabatic conditions. Therefore, a direct time-dependent measurement of both $\dot{T}$ and $s_{ij}D^{pl}_{ij}$ allows the determination of $g(\bar{s}, T_{eff})$. In fact, such a procedure was realized and reported in \cite{99Rittel}. In this work the temperature rise during high strain rates deformation of a glassy polymer was measured. The real-time temperature measurements were obtained by using embedded thermocouples and transient high strain rates of up to $8000 s^{-1}$ were obtained by impact boundary conditions. The plastic power was evaluated by subtracting the elastic power from the total one. Then, the time-dependent temperature record and the known value of $c$ in Eq. (\ref{heat_equationT}) were used to determine $g$ (denoted by $\beta_{diff}$ in \cite{99Rittel}). The most striking result was that $g$ became {\em larger} than unity during fast transients. Moreover, the effect took place during the strain softening part of the stress-strain curve and was more pronounced as the strain rate was increased, all in perfect agreement with the predictions of the proposed theory. Therefore, the results reported in \cite{99Rittel} clearly demonstrate an important physical effect that emerges from our equations.

The discussion that follows Eq. (\ref{heat_equationT}) points to a possible limitation of the HL simulations. The idea is that if strain rates of $\sim 10^4 s^{-1}$ already lead to adiabatic conditions and therefore to a non-negligible temperature rise, then the enormous rates of deformation reached in HL simulations (up to $1\%$ per unit atomic vibration, see Fig. \ref{HL}) must lead to melting of their glass due to the inability of the material to conduct heat sufficiently fast. This physical effect was most probably avoided in the simulations by allowing the numerical thermostat to pump energy from the deforming system at whatever rate needed in order to keep $T$ fixed, effectively using an unrealistic thermal diffusion coefficient. Another manifestation of the same problem can be seen in Fig. \ref{C_s} where the stress can reach values more than
an order of magnitude larger than the yield $s_y$. This observation may question the relevance of the high rates of deformation part of the HL data to understanding plasticity of real materials. However, even if we neglect the very high rates of deformation (and very high stresses) part of the HL data, the remaining part still provides an excellent test for the ideas presented in this work and can be used to isolate novel and important physical aspects of plasticity of amorphous systems. The novel non-monotonic transient effects, which are supported by the available experimental results of \cite{99Rittel} is one of them.

It is important to stress that the non-monotonic behavior of $T_{eff}$ during transients might be very important for understanding strain localization. In \cite{07MLC} it was shown that shear banding initiation takes place only during transients, while the steady states are linearly stable. However, the authors of
\cite{07MLC} used an equation for $T_{eff}$ that precludes a non-monotonic behavior and a strain rate dependent $T^\infty_{eff}$. Redoing the calculation with the $T_{eff}$ dynamics proposed in this work, i.e. Eq. (\ref{heat1}), might shed some new light on this important phenomenon. Moreover, using Eq. (\ref{heat1}) the authors of \cite{BL} found a linear instability in which an expanding circular cavity becomes unstable against the formation of crack-like localized fingers, a result that might be relevant for understanding ``brittle'' versus ``ductile'' behaviors. This clean linear instability was not found for a $T_{eff}$ equation that allows only for a monotonic approach to steady state and see also \cite{07BLPS} for details.

\section{Summary and Discussion}
\label{sum}

In this work an attempt to formulate and rationalize an equation of motion for an effective disorder temperature $T_{eff}$ was made, building on earlier ideas of Langer \cite{04Lan}. The resulting equation, coupled to the predictions of the STZ theory \cite{07BLPa}, suggests a compact deformation theory for athermal amorphous systems. One of the results obtained is that this theoretical structure is consistent with the steady state simulations of \cite{07HL}, for low temperatures. However, the functions $g(s_{ij},T_{eff},T)$ in Eq. (\ref{Teff_dot1}) and $\overline{\C C}(\bar{s})$ in Eq. (\ref{fSTZ}) were not derived from first principles, but were extracted from the numerical data instead. Understanding how to derive these functions from a more fundamental point of view is a major challenge. The present work focused on athermal conditions; understanding how to treat theoretically the contributions of ordinary thermal fluctuations and the interaction of the deforming system with the heat bath for higher temperatures $T$ is yet another challenge. Some steps in this direction were made in \cite{07LM, 07Lan}.

Several recent works proposed alternative $T_{eff}$ dynamics. We have already mentioned that an equation very similar to the one proposed here was suggested in \cite{07HL}. Moreover, the equations used in \cite{07LM, 07BLPS} are fully consistent with the general structure of Eq. (\ref{Teff_dot1}), but with
\begin{equation}
\label{gJIM}
g(s_{ij},T_{eff})=\frac{T_{eff}}{T^\infty_{eff}(\tau_0 \bar{D}^{pl})}
\end{equation}
instead of our Eq. (\ref{g}). Here $T^\infty_{eff}(\tau_0 \bar{D}^{pl})$ is the function plotted in Fig. \ref{HL}. In fact, in \cite{07LM} it was shown that this choice is consistent with the very same HL data analyzed here, implying that neither the choice of Eq. (\ref{g}) nor that of Eq. (\ref{gJIM}) is unique. However, our transient dynamics analysis provides a way to distinguish between the two approaches. Eq. (\ref{gJIM}) immediately implies that $T_{eff}$ reaches its steady state value monotonically, whereas we have demonstrated explicitly that Eq. (\ref{g}) may lead to non-monotonic transients that were observed experimentally in \cite{99Rittel}. This difference, as explained above, may be very important for understanding strain localization that was shown to occur during transients \cite{07LM, 07BL}. More importantly, there is a fundamental difference between the two approaches. When using Eq. (\ref{gJIM}), the function $T^\infty_{eff}(\tau_0 \bar{D}^{pl})$ is an {\em input} to the theory and is not calculated within the theory. However, in our approach, using Eq. (\ref{g}), $T^\infty_{eff}(\tau_0 \bar{D}^{pl})$ is determined by the theory once $\overline{\C C}(\bar{s})$ is known, making the stress dependent rate function $\overline{\C C}(\bar{s})$ a fundamental quantity. Therefore, we still have to understand whether the function $T^\infty_{eff}(\tau_0 \bar{D}^{pl})$ is a fundamental and general quantity, depending only on $\bar{D}^{pl}$, but not on the details of the rearrangements, as was claimed in \cite{07LM}; or that the stress dependent rate function $\overline{\C C}(\bar{s})$ that captures much of the microscopic physics of the irreversible rearrangements is more fundamental, as implied by our approach.

The present work provides further support for the role of the concept of an effective disorder temperature in understanding the irreversible deformation of amorphous systems. In light of these results, based heavily on computer molecular dynamics simulations, one must ask how these ideas are to be tested in laboratory experiments. This raises the question of how to measure $T_{eff}$ in real experiments. In fact, a beautiful example for such a laboratory measurement was give in \cite{05SWM} where the mobility and self-diffusion of tracer particles within a dense granular flow were used to evaluate $T_{eff}$ through an Einstein relation. This measurement is directly related to the work presented here since in \cite{02MK} it was shown explicitly that such an Einstein relation between the self-diffusion coefficient and the mobility coefficient yields a $T_{eff}$ that is consistent with our definition of $T_{eff}$ in Eq. (\ref{Teff_der}). Other possibilities for laboratory measurements of $T_{eff}$ include coupling the system to
a low-frequency harmonic oscillator and measuring its fluctuations \cite{DL} or measuring the barrier
crossing rate of a two-level system weakly coupled to the system of interest, as demonstrated recently in a simulation \cite{07IB}.

An entirely different and new possibility for measuring $T_{eff}$ is related to the quasi-thermodynamic role of $T_{eff}$, as stressed in \cite{07SKLF, 07BLPb}.  In these works $T_{eff}$ was used as a state variable in equations of state for different physical quantities. For example, in \cite{07BLPb} an equation of state relating the mass density $\rho$ to $T_{eff}$ was proposed for simulated amorphous silicon. Once such equations of state become theoretically available one can obtain simple ``$T_{eff}$-thermometers'' by measuring standard macroscopic quantities as a function of the deformation and relating them to $T_{eff}$ through the equation of state \cite{PrivateLanger}. An indication for this possibility was given recently in \cite{07HDJT} where the shear modulus $\mu$ was shown to be a function of the rate of deformation. This quantity can be in principle related to $T^\infty_{eff}(D^{pl})$ once an equation of state of the form $\mu(T_{eff})$ is known. This is yet another future line of investigation emerging from the present work.

{\bf Acknowledgements} I thank T. Haxton and A. Liu for generously sharing with me their numerical data and for useful discussions. I thank J. S. Langer for sharing with me his many ideas on amorphous plasticity and for introducing me with Ref. \cite{07HDJT}. I am grateful to I. Procaccia for continuously and persistently criticizing the concept of the effective disorder temperature. I acknowledge support from the Horowitz Center for Complexity Science and the Lady Davis Trust.

\end{document}